\documentclass{vgtc}                          
\usepackage{siunitx}
\usepackage{hyperref}
\ifpdf
  \pdfoutput=1\relax                   
  \usepackage{graphicx}                
  \DeclareGraphicsExtensions{.pdf,.png,.jpg,.jpeg} 
\else
  \usepackage{graphicx}                
  \DeclareGraphicsExtensions{.eps}     
\fi%

\graphicspath{{figures/}{pictures/}{images/}{./}} 

\usepackage{microtype}                 
\PassOptionsToPackage{warn}{textcomp}  
\usepackage{textcomp}                  
\usepackage{mathptmx}                  
\usepackage{times}                     
\usepackage{cite}                      
\usepackage{tabu}                      
\usepackage{booktabs}                  
\usepackage{multirow}
\usepackage{caption}
\usepackage[final]{changes}  
\usepackage{adjustbox}
\usepackage{etoolbox}
\AtBeginEnvironment{thebibliography}{\interlinepenalty=10000}

\usepackage{tabularx}
\onlineid{1169}

\vgtccategory{Research}

\vgtcinsertpkg

\title{Animation Fidelity in Self-Avatars:\\ Impact on User Performance and Sense of Agency}

\author{Haoran Yun\thanks{e-mail: haoran.yun@upc.edu} %
\and Jose Luis Ponton\thanks{e-mail:jose.luis.ponton@upc.edu} %
\and Carlos Andujar\thanks{e-mail:carlos.andujar@upc.edu} %
\and Nuria Pelechano\thanks{e-mail:nuria.pelechano@upc.edu}}
\affiliation{\scriptsize Universitat Politècnica de Catalunya, Spain}

\teaser{
  \centering
  \includegraphics[width=1\linewidth]{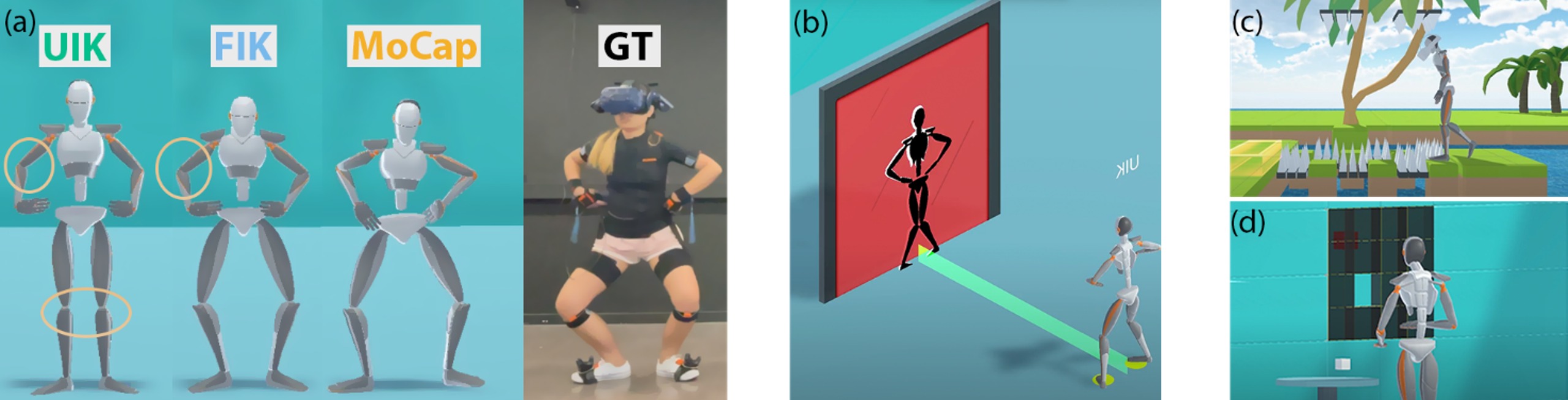}
  \caption{We evaluate self-avatars with three different types of animation fidelity as shown in (a): Unity IK, Final IK and Xsens IMU-based MoCap. We compare them in three tasks involving varied body parts: (b) Copy-pose task, (c) Step-over-spikes task, and (d) Pick-and-place task.}
  \label{fig:teaser}
}

\abstract{The use of self-avatars is gaining popularity thanks to affordable VR headsets. Unfortunately, mainstream VR devices often use a small number of trackers and provide low-accuracy animations. Previous studies have shown that the Sense of Embodiment, and in particular the Sense of Agency, depends on the extent to which the avatar's movements mimic the user's movements. However, few works study such effect for tasks requiring a precise interaction with the environment, i.e., tasks that require accurate manipulation, precise foot stepping, or correct body pose\added{s}. In these cases, users are likely to notice inconsistencies between their self-avatars and their actual pose. 
In this paper\added{,} we study the impact of the animation fidelity of the user avatar on a variety of tasks that focus on arm movement, leg movement and body posture. We compare three different animation techniques: two of them using Inverse Kinematics to reconstruct the pose from sparse input (6 trackers), and a third one using a professional motion capture system with 17 inertial sensors. We evaluate these animation techniques both quantitatively (completion time\deleted{s}, unintentional collisions, pose accuracy) and qualitatively (Sense of Embodiment). 
Our results show that the animation quality affects the Sense of Embodiment. Inertial-based MoCap performs significantly better in mimicking body poses. Surprisingly, IK-based solutions using fewer sensors outperformed MoCap in tasks requiring accurate positioning, which we attribute to the \added{higher latency and the} positional drift that causes errors at the end-effectors, which are more noticeable in contact areas such as the feet.%
} 

\CCScatlist{
  \CCScatTwelve{Computing methodologies}{Computer graphics}{Graphics systems and interfaces}{Virtual reality,Perception};
  \CCScatTwelve{Computing methodologies}{Computer graphics}{Animation}{Motion capture}
}

\begin{document}



\maketitle

\section{Introduction} 

Virtual reality headsets allow us to immerse ourselves in highly realistic digital worlds. A fundamental aspect of feeling present in these virtual environments is to have a virtual representation of our own body, known as the user's self-avatar. Ideally, avatars should be animated in a way that allows users to achieve natural behaviors and interactions in the virtual environment, as well as to use non-verbal communication with others.
\added{Self-avatar is the virtual representation of oneself and should be distinguished from other people's avatars as they have different requirements. Aspects such as latency or end-effectors and pose accuracy are more crucial for perceiving one's own avatar than for others.}


Unfortunately, the limited number of trackers in consumer-grade devices severely \replaced{restricts}{limits} the quality of the self-avatar's movements. Most applications limit the representation to floating upper bodies (no legs) with floating hands/globes/tools, sometimes with the arms animated with Inverse Kinematics (IK), by using the tracking data from the HMD and the hand-held controllers. Only a few applications offer a full-body representation. However, due to the lack of trackers, legs are typically animated by playing cyclic time-warped animations. With these solutions, users may notice inconsistencies between their movements perceived via proprioception and those of the self-avatar.

Previous work has demonstrated the importance of having a self-avatar that moves in sync with the user \cite{fribourg_AvatarSenseEmbodiment_2020,galvandebarba_PlausibilityVirtualBody_2020,toothman_ImpactAvatarTracking_2019a}. If we focus on the overall movement without further interaction with the virtual world, current animation techniques based on IK from sparse tracking data may suffice. However, if accurate body poses and positioning of end-effectors matter, then artifacts that affect user performance and the Sense of Agency \deleted{(SoA)} may pop up. For example, consider the task of building some assembly by holding pieces and putting them in specific locations. In that case, hand-eye coordination is crucial, as is the accuracy of the overall pose, to prevent parts of the arm/body from colliding with other pieces. Another example is moving through a room full of obstacles, where accurate foot positioning is also crucial. Finally, correct body poses also matter in the case of learning to dance or practicing yoga by mimicking an instructor \cite{fribourg_AvatarSenseEmbodiment_2020}. 

Given that high-quality motion capture is difficult to achieve with sparse input data, we are interested in studying how animation fidelity affects user performance and embodiment. By animation fidelity, we refer to the quality of the animations in terms of accurately following the user poses as well as the correct absolute positioning of the body parts. 
More specifically, we evaluate interactions with the virtual world that need pose and/or positional accuracy. We evaluate embodiment with a perceptual study, in which our main focus is on the Sense of Agency due to its relationship with animation fidelity. Furthermore, we study the effect of the quality of the interaction with the virtual world on user performance by measuring completion time\deleted{s} and unintentional collisions. We focus on two popular methods based on Inverse Kinematics from sparse input data (6 trackers): UnityIK\footnote{https://docs.unity3d.com/Manual/InverseKinematics.html} and FinalIK\footnote{http://root-motion.com/}, and one motion capture system based on a large number (17) of inertial sensors: Xsens Awinda\footnote{https://www.xsens.com/products/mtw-awinda}. 


Our results suggest that animation fidelity affects the Sense of Embodiment and user performance. We found that a straightforward IK solution, such as Unity IK, decreases the Sense of Embodiment when compared to high-quality IK and MoCap solutions. However, when interacting with the environment, having \added{lower latency and} minimal end-effector positional error may be more important than synthesizing high-quality poses suffering from positional drift.

The main contributions of this paper are:
\begin{itemize}
    \vspace{-0.1cm}\item To the best of our knowledge, this is the first study to compare an IMU-based full-body motion capture system to IK solutions for animating self-avatars in VR during tasks that require accurate positioning of end-effectors and body postures.
    \vspace{-0.1cm}\item We study the relationship between animation fidelity on user task performance and \added{the} Sense of Agency to improve future research on VR avatar animation from sparse data.
\end{itemize}

\section{Related Work}
\subsection{Self-avatars and animation fidelity}
A self-avatar is a virtual representation of one's \added{own} body from a first-person view of the virtual environment \added{(VE)}. Previous studies have shown that full-body self-avatars are beneficial in various tasks\added{,} such as egocentric distance estimation, spatial reasoning tasks, and collision avoidance \cite{ries_Analyzingeffectvirtual_2009,pan_AvatarTypeAffects_2019,pan_HowFootTracking_2019}. For instance, compared to not having an avatar, users with a full-body realistic avatar collide less frequently with the \replaced{VE}{virtual environment} \cite{pan_HowFootTracking_2019}. Similarly, Ogawa et al. \cite{ogawa_YouFeelPassing_2020} demonstrated that users would be less likely to walk through the virtual walls if equipped with a full-body representation compared to a hands-only representation. In social settings, using full-body self-avatars would enhance social presence and communication efficiency  \cite{yoon_EffectAvatarAppearance_2019,aseeri_InfluenceAvatarRepresentation_2021}.

Animation fidelity is a crucial component of self-avatars. Unlike visual fidelity, which addresses the appearance of avatars and has been extensively studied  \cite{goncalves_Evaluationimpactdifferent_2022,ogawa_YouFeelPassing_2020,gao_EffectsAvatarVisibility_2020,choi_EffectsLocomotionStyle_2020a}, animation fidelity focuses on how accurately and synchronized the self-avatar mimics users' real-world movements. We could use avatars with the highest visual fidelity (with a realistic full-body self-avatar), but low animation fidelity if the body poses are not well mimicked, not in sync with the user, or have errors in the positioning of end-effectors. These avatars are unlikely to induce the user's feeling of owning or being in control of the virtual body \cite{kilteni_myfakebody_2015,Waltemate}. These kind\added{s} of problems may be introduced by the tracking system or by the methods used to capture and animate the self-avatar. 

Inverse Kinematic (IK) solvers can be used with sparse input from VR devices to calculate joint angles of the articulated human model. Some frameworks are available to animate full-body avatars from six trackers (HMD, two hand-held controllers and three Vive trackers) \cite{ponton2022avatargo, oliva2022quickvr,zeng_PEDLSnovelmethod_2022}. Parger et al. \cite{parger_HumanUpperBodyInverse_2018} proposed an intuitive IK solution for self-avatar's upper-body animation with one HMD and two controllers. Their IK solver outperformed an optical MoCap system in terms of lower latency and accurate pose reconstruction. The reduced Jacobian IK solver proposed by Caserman et al. \cite{caserman_Realtimebodytracking_2019} can smoothly and rapidly animate full-body self-avatars with HTC Vive trackers. 

Recently, researchers have shown an increasing interest in data-driven methods to reconstruct full-body animation for avatars from VR devices. For instance, Winker et al. \cite{winkler2022questsim} proposed a reinforcement learning framework with physical-based simulation to achieve real-time full-body animation. Ponton et al. \cite{ponton2022mmvr} combined body orientation prediction, motion matching and IK to synthesize plausible full-body motion with accurate hand placement. Jiang et al. \cite{jiang_AvatarPoserArticulatedFullBody_2022a} used a transformer model to estimate the full-body motion. Other researchers have looked at using a sparse set of wearable IMUs to estimate full-body motion. These methods could be integrated into self-avatars in VR because of the built-in IMUs on VR headsets, controllers and trackers. For example, Huang et al. \cite{DIP:SIGGRAPHAsia:2018} used a bi-directional RNN to reconstruct a full-body human pose from six wearable IMUs attached to the head, arms, pelvis, and knees. Yi et al. \cite{yi_TransPoserealtime3D_2021} \replaced{took}{takes} the same input, but \replaced{generated}{generates} both accurate pose and precise global translation. \replaced{More}{Most} recently, Jiang et al. \cite{jiang_TransformerInertialPoser_2022} not only accurately \replaced{estimated}{estimates} the full-body motion but also \replaced{handled}{handles} the joint and global position drift that most IMU systems suffer from. 

While there is an extensive body of research proposing new animation methods to improve animation fidelity for avatars, little interest has been given to how the animation fidelity of self-avatars impacts user performance, perception and behavior in a \replaced{VE}{virtual environment}. Fribourg et al. \cite{fribourg_AvatarSenseEmbodiment_2020} showed that users preferred to improve animation features when asked to choose among appearance, control (animation) and point of view, to improve the \added{Sense of Embodiment} \added{(}SoE\added{)}. \added{In their work, participants preferred mocap based on Xsens over FinalIK. However, their input to the IK system was the joints positions from the Mocap system, and thus the problems with incorrect end-effector positioning and latency were carried on to the IK condition.}

Galvan et al. \cite{galvandebarba_PlausibilityVirtualBody_2020} adapted the same methodology to examine the effect of animation fidelity of different body parts. Participants were first exposed to optimal animation fidelity (53-marker optical motion capture). Then, they started with minimal animation fidelity and repeatedly chose one body part to improve until they felt the same level of the SoE as with the optimal configuration. They found users felt the same level of the SoE with an IK solution with eight trackers than with the 53-marker optical motion capture system. Their work also found that the unnatural animation of the full body caused disturbing feelings for users when separately animating the upper body and lower body with different fidelity. Thus, our work focuses on full-body animation instead of body parts to avoid breaking the user's presence. Eubanks et al. \cite{eubanks_EffectsBodyTracking_2020} explored the impact of the tracking fidelity (number of trackers) on a full-body avatar animated by an IK solver. They found that a high number of trackers could improve the SoE. However, animation fidelity is not only about tracking fidelity, but also about the animation techniques underneath. Our study thus \added{compares} not only \deleted{compares} systems with different numbers of trackers, but also different animation techniques: IK and IMU-based motion capture. 

\subsection{Sense of Agency}\label{subsec2}
\added{The} Sense of Agency (SoA) has been characterized in various ways in different contexts because of its interdisciplinarity property. From the point of view of psychology, \added{the} SoA refers to the feeling that \textit{I am the one causing or generating an action} \cite{david2008sense}. In the field of VR, \added{the} SoA is the feeling of being the agent who conducts the motions of an avatar. It results from synchronizing one's real-world movements with virtual body motions. 

\replaced{The Sense of Agency}{SoA} is a crucial subcomponent of \replaced{the Sense of Embodiment}{SoE}. According to Kilteni et al. \cite{kilteni_SenseEmbodimentVirtual_2012}, \added{the} SoE consists of three subcomponents: \replaced{the Sense of Agency}{SoA}, \added{the} Sense of Self-Location (SoSL), and \added{the} Sense of Body Ownership (SoBO). Many studies have studied the impact of single or multiple factors, including avatars' appearance, visibility and tracking fidelity, on the SoE. 
Fribourg et al. \cite{fribourg_AvatarSenseEmbodiment_2020} explored the relative contributions of the control factor (i.e.\deleted{,} animation fidelity), appearance and point of view that contribute to the SoE. Results showed that control and the point of view were preferred when people had to choose among the three factors to improve their SoE. Recent studies showed that low-quality tracking, which directly impacts the animation of self-avatar, can decrease the embodiment \cite{toothman_ImpactAvatarTracking_2019a,eubanks_EffectsBodyTracking_2020}. These findings analyzed the effect of the SoE, which is directly or implicitly related to animation. However, there is still a gap in how the animation fidelity directly impacts the SoE, specifically the subcomponent SoA.

The synchronicity of visuomotor correlation can induce the \replaced{SoA}{Sense of Agency}\added{,} while discrepancies can decrease it. Kollias et al. \cite{koilias_EffectsMotionArtifacts_2019} simulated different kinds of motion artifacts that may occur in a real-time motion \replaced{capture}{capturing} system\replaced{. They}{and} examined the effect of these artifacts on \added{the} SoE, specifically on \added{the} SoA. Results showed that the artifacts negatively \replaced{affected }{affect} the SoA, but not \added{the} SoBO.


Studies regarding \added{the} \replaced{SoA}{Sense of Agency} mainly focus on subjective perception with questionnaires and objective brain \replaced{activity}{activities} measurements such as fMRI and EEG. As suggested by Kilteni et al. \cite{kilteni_SenseEmbodimentVirtual_2012}, the motor performance of VR users should be positively correlated with the \replaced{SoA}{Sense of Agency}, under the assumption that a fine-controlled virtual body performs motor tasks more successfully. Therefore, the users' motor performance in VR could be used to measure \added{the} SoA objectively. \replaced{Our study}{In our study, we} measured task performance in terms of unintentional collisions between the self-avatar and the virtual obstacles. We believe that the number of collisions and their duration could bring insights into human motor performance in 3D space. High animation fidelity means precise control of self-avatars which can perform better in motor tasks. Therefore, we expected to observe the impact of animation fidelity on collisions, completion time\deleted{s}\added{,} and \added{the} correctness of the body poses.


\section{Animation fidelity study}
This study aims to assess the importance of animation fidelity on the users' performance and the \replaced{SoE}{Sense of Embodiment} when performing a set of tasks that require careful positioning and/or accurate poses. We want to study the importance of \deleted{having} the virtual body correctly mimicking the user\added{'s} movements \replaced{as well as}{but also} the impact \replaced{of accurate end-effector positioning}{on having accurate positioning of end-effectors}.
\deleted{More specifically, we want to answer the following questions: do animations driven by Inverse Kinematics exhibit similar or lower levels of embodiment than animations driven by an inertial motion capture system? To what extent does the animation fidelity of a self-avatar affect users' performance during tasks that require interaction with the virtual environments? To what extent does the animation fidelity of a self-avatar affect users' perceived the \replaced{SoE}{Sense of Embodiment} and its subcomponents?}






\subsection{Experimental conditions}

\begin{figure}[tb]
 \centering 
 \includegraphics[width=0.8\columnwidth]{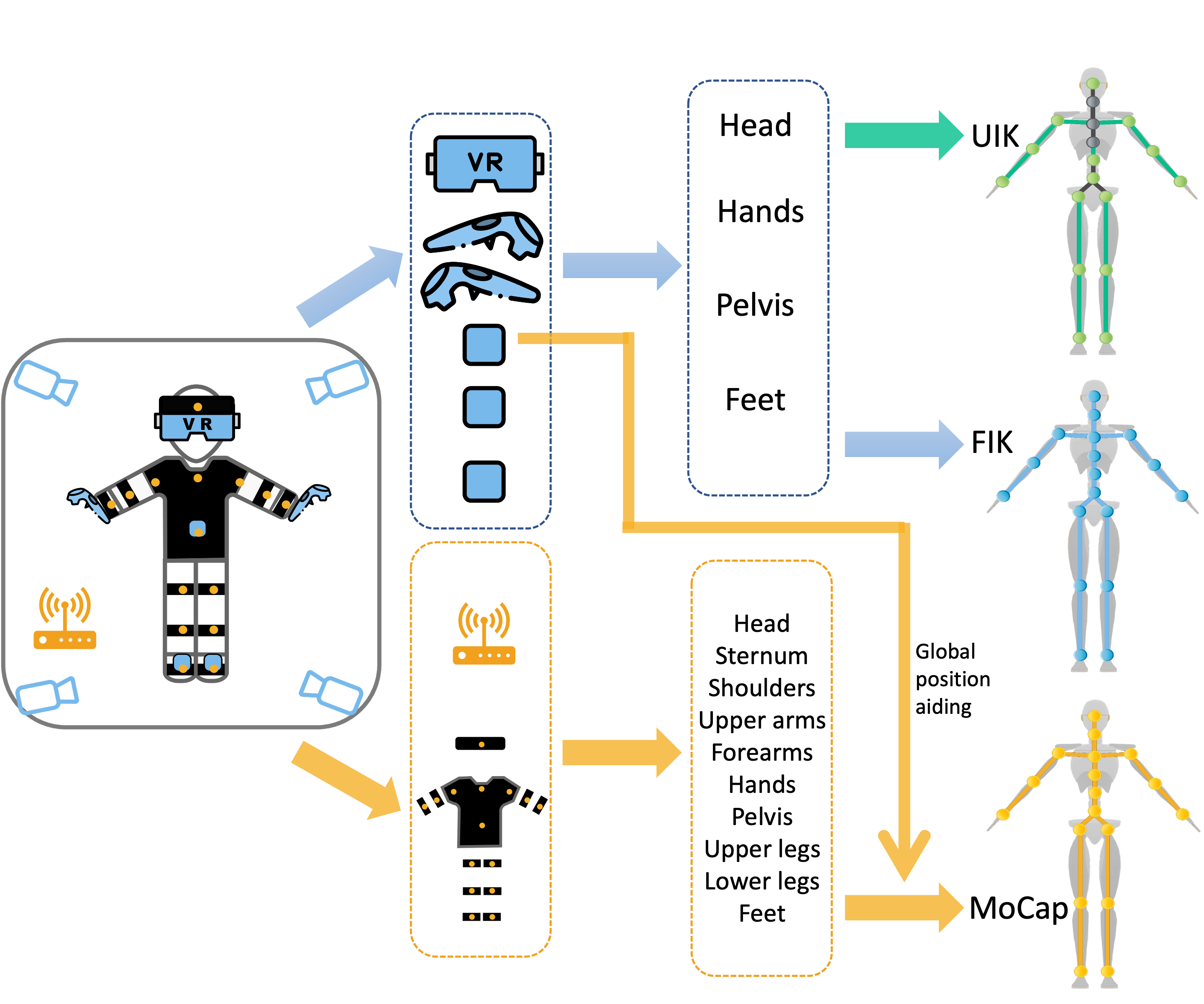}
 \caption{Equipment and conditions. For the experiment, participants were \added{simultaneously} equipped with two sets of tracking devices: VR devices (\deleted{HTC Vive} HMD, controllers and three trackers); and the trackers from the Xsens Awinda \replaced{mocap}{motion capture} system. The tracked body parts are shown in the figure. Different IK solvers were applied to animate the avatar using the VR tracking devices. }
 \label{fig:flowchart}
\end{figure}

In this study, we adopted a within-subject experiment design with one independent variable: the animation fidelity for the virtual avatar. We designed three conditions for the animation fidelity variable: Unity Inverse Kinematics (UIK), FinalIK (FIK) and motion capture with Xsens (MoCap). These techniques provide different levels of animation quality in terms of end-effector positioning (more accurate in UIK and FIK since hand-held controllers and trackers provide accurate absolute positioning)\replaced{,}{ and} pose angles (more accurate in MoCap thanks to a larger number of sensors)\added{, and latency (higher for MoCap)}. The first two conditions differ on the IK solvers\added{,} while both use sparse tracking data from consumer-level VR devices. The last condition, MoCap, \replaced{uses}{used} tracking data from a professional motion capture system with 17 IMU sensors. \replaced{Fig.}{Figure}~\ref{fig:flowchart} illustrates the equipment used for tracking in the three conditions.
The three conditions used have been implemented as follows (see accompanying video):

\noindent
\textbf{UIK}: This condition uses Unity 2020.3 game engine's built-in IK solver for animating the avatar's limbs (2-segment kinematic chains). It is important to note that it does not consider the full-body pose when solving the IK. Instead, it independently computes each limb's joints based on \replaced{one}{ the} target end-effector\deleted{s}. To further improve the overall body pose, \deleted{we have included}forward kinematics (FK) is included to animate two joints: head and spine, so that the self-avatar can lean forwards and sideways. IK and FK together generate a full-body animation for the avatar from the tracking data in the HMD, the hand-held controllers and three additional trackers located on the pelvis and the feet.

\noindent
\textbf{FIK}: This condition uses \added{the} VRIK \added{solver} from RootMotion's FinalIK package, which combines analytic and heuristic IK solvers for generating the full-body avatar animation. \replaced{With the same input, FIK produces higher-quality results than UIK because each limb is not solved independently from one end-effector, but rather from an analysis on the user pose from several end-effectors  \cite{wagnerberger_Inversekinematicsfullbody_2021}. For instance, the spine is solved considering the position of the HMD and two hand-held controllers, and the elbows use the position of the hands relative to the chest joint to determine the orientation. The only exception are the legs, which are solved independently but using a 3-joint dual-pass trigonometric solver (first solve the knee and then the ankle).}{This
condition uses the same tracking data as UIK. However, the IK solver considers limbs and spines as 4-segment kinematic chains to generate nature-looking poses efficiently. It achieves more natural poses than UIK since it takes into account the full body when solving the IK.}


\noindent
    \textbf{MoCap}:  \replaced{The Xsens Awinda system receives acceleration, angular velocity and magnetic field data from 17 body-worn IMUs, processes the data with Strap-down Integration and Kalman filtering, and then outputs the rotations of the joints of the avatar, which are streamed to Unity via UDP; these processing steps increase the latency with respect to the previous conditions.} {Xsens Awinda is an inertial motion capture system that uses 17 IMU sensors providing the rotations of all major joints for the avatar.} IMUs suffer from a positional drift over time, that might break the Sense of Self-location. To enforce the correct location of the avatar with the user, we use the \added{pelvis} tracker \deleted{located at the pelvis} to position the avatar in the \replaced{VE}{virtual environment}. However, this does not guarantee accurate positioning of the end-effectors and can suffer from foot sliding. Foot lock is applied to reduce the foot sliding of the \deleted{given pose by} Xsens \added{pose} when the foot touches the floor. Once the foot is locked, we \deleted{also} store the position of the HTC tracker, which we will use as a reference to detect whether the user is moving the foot. In the following frames, if the distance between the current HTC tracker and its initial position is larger than a relatively small threshold (1\,cm), we unlock the foot; otherwise, it will noticeably modify the leg pose. Note that we are locking the foot at the position given by Xsens (thus, it may contain positional error); we only use the HTC tracker to detect whether the user's foot remains on the ground. 

Each participant performed the same three tasks for each condition\replaced{. Conditions were counterbalanced between participants using a Balanced Latin Square, which ensures each condition precedes and follows every other condition an equal number of times \cite{balanced_latin_square}.}{,with the three conditions being counterbalanced between participants using a Balanced Latin Square.}

\begin{figure*}[tb]
 \centering 
 \includegraphics[width=0.9\linewidth]{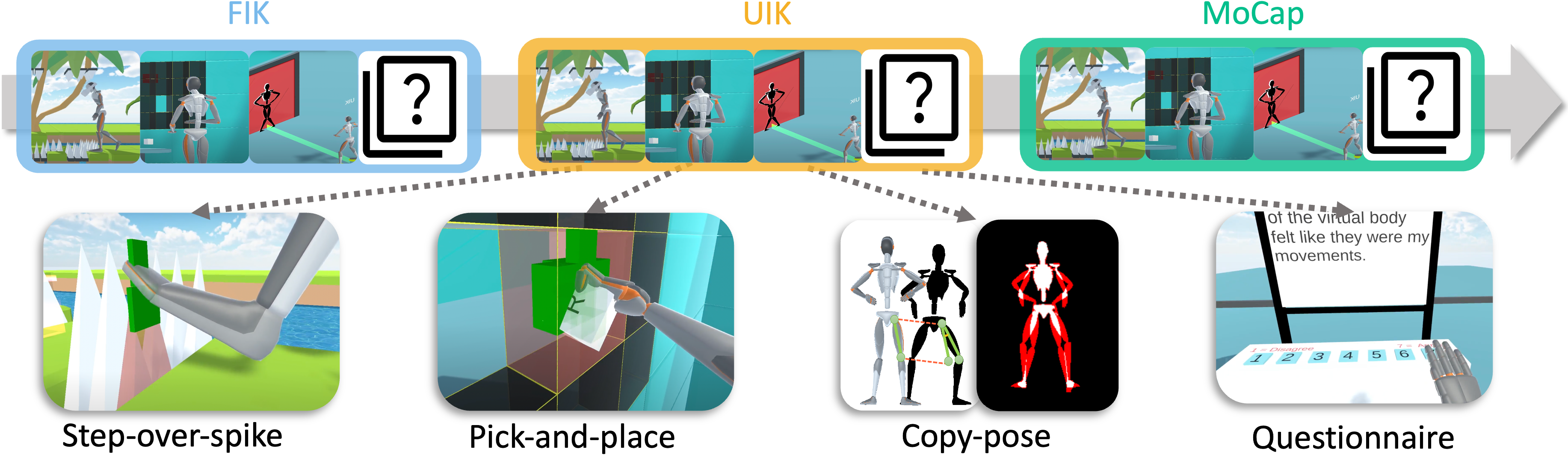}
 \caption{Timeline for one participant (top) with details on the tasks (bottom). The participants were asked to complete three consecutive tasks followed by the questionnaire for the first condition, and then repeat the procedure with the other two conditions. During the first two tasks, the volume of the small colliders were recorded (green), and users had visual feedback from the obstacles (in red) every time a collision occurred. For the copy-pose task, the pose-related metrics were calculated. Questions and buttons for answering were displayed on a whiteboard in VR.}
 \label{fig:timeline}
\end{figure*}

\subsection{Tasks}
Prior studies have shown that the type of actions users perform in a \replaced{VE}{virtual environment} influence\added{s} users' perception \cite{fribourg_AvatarSenseEmbodiment_2020,tuveri_FitmersiveGamesFitness_2016}. For instance, when picking \replaced{up nearby objects}{nearby objects up}, people would pay more attention to the upper body while \added{their} ignoring surroundings \cite{cmentowski_EffectsTaskType_2021}. Similarly, walking in a room with obstacles on the floor would draw people's attention to objects and lower body parts to plan the future path and avoid collisions \cite{stein_EyeTrackingbasedLSTM_2022}. We thus designed three tasks that cover a wide range of common actions in VR games and applications, while each task \replaced{focused}{focuses} on a different interaction pattern between the virtual body and \replaced{the VE}{virtual environment} (see \replaced{Fig.}{Figure} \ref{fig:timeline} and accompanying video). 
\begin{itemize}
    \vspace{-0.1cm}\item \textit{Step-over-spikes task} focuses on direct interaction between the lower body and the VE. It consists of walking and lifting the knees to avoid colliding with spike-like obstacles while stepping on small platforms.
    \vspace{-0.1cm}\item \textit{Pick-and-place task} focuses on direct interaction between the upper body and the VE. It consists of picking up objects\deleted{,} and then placing them at specific target locations while avoiding collisions between the arm and nearby obstacles.
    \vspace{-0.1cm}\item \textit{Copy-pose task} involves only non-direct interactions between the virtual body and the VE. More specifically, we focus on the overall pose of the self-avatar without caring about the exact global positioning of the avatar. For this task\added{,} we show a 2D projection of an avatar in a certain pose on a virtual screen, and then the user is asked to mimic the pose as accurately as possible. The design is inspired by OhShape\footnote{https://ohshapevr.com/}.
\end{itemize}

One task block \replaced{consisted}{consists} of the following three sequential tasks\replaced{, which were presented in the following order}{in a predefined order}: (1) step-over-spikes task; (2) pick-and-place task; (3) copy-pose task. Each task \replaced{consisted}{consists} of ten trials separated by five seconds of rest. \added{We decided to use this task order to guarantee that the last task before each SoE questionnaire (see below) equally involved the upper and lower limbs.} 
Participants were requested to complete the entire task block for each of the three conditions on a recurrent basis. 

\subsection{Apparatus}
The experiments were conducted in an acoustically-isolated laboratory room with \added{a} 3\,m x 6\,m space\deleted{surrounded by four HTC Vive base stations}. The VE was developed with Unity 2020.3\, LTS and run on a PC equipped with a CPU Intel Core i7-10700K, a GPU Nvidia GeForce RTX 3070 and 32\,GB of RAM. We used an HTC Vive Pro HMD with 1440 × 1600 pixels per eye, 110$^{\circ}$ field of view and 90 Hz refresh rate. Three 6-DoF HTC Vive trackers 3.0 were used for tracking \replaced{the pelvis and feet}{three body parts: pelvis, left and right feet}. Two HTC Vive controllers were held in both hands of the participants. We installed four SteamVR Base Station 2.0 in each corner of the room to minimize line-of-sight occlusions. 

\added{We employed the well-established frame counting approach \cite{caserman_Realtimebodytracking_2019,steed_simplemethodestimating_2008} 
to determine the latency of the tracking system and the animation techniques used in our experiment. One person was equipped with all tracking devices and repeatedly moved one arm up and down. We used a high-speed 240fps camera to record both the person and a monitor showing the VE. 
The mean latency from the physical controller to the animated virtual hand was 32 ms for UIK and 33 ms for FIK. These latencies include the SteamVR tracking system, IK solver computation and rendering. For MoCap, the mean latency 
was 91 ms, which was notably higher than the other two conditions. The MoCap latency includes the IMU-to-Xsens software latency ($\sim$30 ms}
\footnote{https://base.xsens.com/s/article/MVN-Hardware-Overview}\added{)   
\cite{paulich2018xsens}, motion processing in Xsens software, network communication, motion data unpacking in Unity ($\sim$5 ms), and rendering. }

\subsection{Procedure}

A total of 26 participants took part in the experiment (22 male, 4 female, aged 19-40, M = 22.4, SD = 5.5) but one participant's data was discarded due to \added{a} calibration failure.

Upon arriving at the experiment room, participants were instructed to read the information sheet and complete the consent form and a demographic survey detailing their age, gaming and VR experience. We introduced their body measurements to the Xsens software in order to obtain a scaled avatar matching the user's dimensions. Then we placed the 17 wireless trackers on the body of the participant, with the help of a t-shirt and a set of straps. Participants were asked to walk a few meters to calibrate the IMU-based motion capture suit. \replaced{The calibration was repeated until the Xsens software (MVN Animate Pro) rated it as a "Good" (among "Good", "Acceptable" and "Poor"). The experimenter also validated visually that the subject's pose closely matched that of the avatar.}{The experimenter carefully inspected the calibration results to ensure that the motion capture device functioned properly with the scaled avatar.} Next\added{,} participants were equipped with an HTC Vive HMD, two hand-held controllers and three Vive trackers placed on the pelvis and both feet. They were asked to stand in a T-pose to complete the calibration of the HTC trackers for the IK solvers. During the experiment, participants were equipped with both tracking systems at all times. This ensured that they could not guess what system was being used for each condition. Before each task\added{,} participants watched a tutorial video (two minutes in total) that demonstrated how to perform the task. 

\subsection{Measurements}
\begin{table}
\scriptsize
    \centering
    \begin{tabu}{X}
    \toprule 
    \hline \addlinespace[0.1cm] \textit{\textbf{Agency} - Scoring: (AG1 + AG2 + AG3 + AG4 + AG5 + AG6 + AG7) / 7} \\
    \textbf{AG1} The movements of the virtual body felt like they were my movements.\\
    \textbf{AG2} I felt the virtual arms moved as my own arms.\\
    \textbf{AG3} I felt the virtual elbows were in the same position as my own elbows.\\
    \textbf{AG4} I felt the virtual hands were in the same position as my own hands.\\
    \textbf{AG5} I felt the virtual legs moved as my own legs.\\
    \textbf{AG6} I felt the virtual knees were in the same position as my own knees.\\
    \textbf{AG7} I found it easy to control the virtual body pose to complete the exercises.\\ \addlinespace[0.1cm] 

    \hline \addlinespace[0.1cm] \textit{\textbf{Ownership} - Scoring: (OW1 + OW2 + OW3) / 3}\\
    \textbf{OW1} It felt like the virtual body was my body.\\
    \textbf{OW2} It felt like the virtual body parts were my body parts.\\
    \textbf{OW3} It felt like the virtual body belonged to me.\\\addlinespace[0.1cm]

    \hline\addlinespace[0.1cm] \textit{\textbf{Change} - Scoring: (\deleted{flipped(}CH1\deleted{)} + \deleted{flipped(}CH2\deleted{)}) / 2}\\
    \textbf{CH1} I felt like the form or appearance of my own body had changed.\\
    \textbf{CH2} I felt like the size (height) of my own body had changed.\\ 
    \bottomrule\addlinespace[0.05cm]
    \end{tabu}
     \caption{Questionnaire content. The scores are on \added{a} 7-Likert scale (1 = \added{completely} disagree, 7 = \added{completely} agree). \deleted{The answer scores for CH1 and CH2 are to align them with other scores so that higher means better. For example, if one gives x for CH1, then the flipped score will be (7 - x + 1 ).}}
    \label{tab:questions}

\end{table}

The step-over-spikes task challenges the participants' lower-body motion so that we can quantitatively assess the effect of animation fidelity on the interaction between the lower body and the VE. Similarly, the pick-and-place task is intended to assess the impact of animation fidelity on \added{the} interaction between the upper body and the VE. 
To evaluate these two tasks\added{,} we took measurements regarding collisions and completion time. More specifically, we recorded: the total collision volume ($V_{c}$), the collision duration ($T_{c}$), the number of collisions ($N_{c}$), as well as the task completion time ($T_{task}$). This data was converted into more intuitive metrics as follows:
\begin{itemize}
    
    \vspace{-0.1cm}\item Volume per collision $v = V_{c} / N_{c}$. It reflects how deep the avatar penetrated the obstacle during each collision, on average.
    \vspace{-0.1cm}\item Duration per collision $t = T_{c} / N_{c}$. It measures the average penetration time of the avatar into obstacles and how quickly participants corrected the collision when it occurred.
    \vspace{-0.1cm}\item Collision frequency $f = N_{c} / T_{task}$. It reflects how often the avatar collides with obstacles while performing the task. It is specified as the number of collision\added{s} per second.
\end{itemize}
With these metrics, we investigated the relationship between the animation fidelity and \replaced{the virtual body-obstacle collisions}{the collisions between the participants' virtual bodies and the obstacles}. To accurately capture the volume and duration of collisions, a set of invisible small cubic colliders ($V_{collider} = 8$\,cm$^{3}$) were used to match the shape of each obstacle.

The goal of the copy-pose task is different from the other two. It evaluates the correctness of the static pose of the avatar when there are no hard constraints for the avatar's end-effectors positions (i.e. no \deleted{direct interaction} \added{contact points} between the avatar and the VE). Thus, three pose-related metrics were used to assess the accuracy of users' poses: 
\begin{itemize}
    \vspace{-0.1cm}\item Jaccard Distance  $JD = 1 - \frac{G \cap U}{G \cup U}$ (see \replaced{Fig.}{Figure}~\ref{fig:timeline}). It measures the non-overlap area of the intersection between the 2D projection $G$ of an example avatar over a plane and the 2D projection $U$ of the avatar controlled by the user, divided by the union of the two projections.
   \vspace{-0.1cm} \item Mean per segment angle error ($MPSAE$) is defined as: ${MPSAE = \frac{1}{\mathrm{\|S\|}} \sum_{\hat{s}}^{\mathrm{\|S\|}} \arccos \left( \hat{s}^* \cdot \hat{s} \right)}$, where $\mathrm{S}$ is the set of segments of the skeleton, $\hat{s}$ is the unit vector representing the direction of a segment $s$, and $\hat{s}^*$ is the direction of the segment in the given pose.
    \vspace{-0.1cm}\item Mean per part angle error $MPPAE$ is like $MPSAE$ but only considers one part of the body such as the spine or the limbs corresponding to arms and legs. 
\end{itemize}
Participants could only observe the target poses as a 2D projection on a virtual wall that was located in front of them. Therefore, the metrics used in this task \replaced{were}{are} all calculated based on the same 2D projection in the XY plane. For the Jaccard Distance, the lack of overlap between the two projections must not be \replaced{a result of}{due to} the user position being slightly offset with respect to the observed shape. Consequently, we iteratively applied translations in the \replaced{2D}{two-dimensional} space to maximize the overlap between the two shapes before computing \deleted{the} $JD$. For $MPPAE$, body segments of the avatar were grouped into three body parts: arms, legs and spine. This separation \replaced{allowed}{allows} us to study animation fidelity's impact individually on different body parts.

At the end of each block of tasks, participants completed a questionnaire (Table \ref{tab:questions}) which was adapted from a standard questionnaire from Virtual Embodiment Questionnaire (VEQ) \cite{roth_ConstructionVirtualEmbodiment_2020}. \replaced{The embodiment}{Embodiment} was measured through three main aspects: \textit{agency}, \textit{ownership} and \textit{change}. \textit{Agency} measures the sense of control, \textit{ownership} measures the sense of owning the virtual body as if it is one's own real body, and \textit{change} measures to what extent one feels the virtual body scheme differs in size from one's own real body. 

The VEQ does not assess self-location discrepancies since it is not the goal of typical VR applications to produce such effects~\cite{roth_ConstructionVirtualEmbodiment_2020}. In our experiment, the use of the pelvis tracker guaranteed \added{a} \deleted{an overall} correct placement of the self-avatar. The appearance and size of the avatar were kept the same through all conditions to guarantee that the only perceived differences would come from the changes in animation fidelity. \added{Questions about \textit{change} in VEQ are typically studied in the context of body swap experiments that manipulate avatars' body size, sex, race, etc. \cite{Wolf_2021, Kocur_2021, Wirth_2021}. However, with the avatar's height and body proportions consistent with the user's physical body, \textit{change} is not expected to be an influencing factor in our study.}   

\added{The goal of the embodiment questionnaire was to gather the global experience after running the three tasks, so that it would gather both the importance of correct end-effector positioning and the accuracy of the body pose. 
We decided against asking the 15 questions after each task to avoid doing the experiment too long because it could introduce a biased source.}


\subsection{Hypotheses}
We hypothesize that better animation fidelity would lead to better performance in terms of reducing the \replaced{number}{amount} of collisions, and also their volume and duration. 
\added{Although our conditions had varied trade-offs in terms of the different components of animation fidelity (pose accuracy vs end-effector accuracy, as well as latency), we expected}
\deleted{Therefore we should expect} the highest performance for \replaced{the full-body IMU-based motion capture system}{Mocap}, followed by \replaced{IK methods with VR devices as input}{FIK and then UIK}. Similarly we would expect \replaced{the full-body IMU-based motion capture system}{Mocap} to outperform the IK solution when copying body poses given its higher number of sensors allowing for a more accurate capturing of the user pose. Finally we expect\added{ed} animation fidelity to affect the \replaced{SoE}{Sense of Embodiment} of the user.
%
%
%
%
%
\replaced{Therefore, }{More precisely,} our hypotheses are:

\begin{description}
\vspace{-0.1cm}\item[H1] Animation fidelity impacts performance of the user in \added{step-over-spikes and pick-and-place} (tasks that require precise interaction with the environment)\added{, in terms of unintended collisions and completion time}. 
\vspace{-0.1cm}\item[H2] Animation fidelity impacts performance in \replaced{copy-pose task}{tasks}, which requires accuracy in the body pose. 
\vspace{-0.1cm}\item[H3] Animation fidelity affects the \replaced{SoE}{Sense of Embodiment}.
\end{description}


%
\section{Results}
\added{In this section we summarize the results of our experiment. The complete list of statistical significance and post-hoc tests values can be found in Table \ref{tab:statistic}.}

\begin{table*}[tb]
  \scriptsize%
	\centering%
 \adjustbox{max width=0.772\textwidth}{
  \begin{tabu} {%
	l%
	*{1}{l}%
	*{1}{X}%
	}
    
\textbf{Metric} & \textbf{Test} & \textbf{Post-hoc}\\
  \toprule
    \multicolumn{3}{l}{\textbf{\textit{Step-over-spike Task}}}\\\hline
    \multirow{ 2}{*}{Volume Per Collision ($v$)} &  Friedman test & Wilcoxon test with Bonferroni adjustment\\
                         &$\chi^2(2) = 11.80 , p = .003, W = .235 $ & MoCap $>$ UIK ( $p  = .014 \added{,r = .552}$), MoCap $>$ FIK ($p=.009\added{, r= .573}$)\\ \hline \addlinespace[0.15cm] 
    \multirow{ 2}{*}{Duration Per Collision ($t$)} &  Friedman test & -\\
                         &$\chi^2(2) = 4.16, p = .125(ns), W = .083$ & -\\\hline \addlinespace[0.15cm] 
    \multirow{ 2}{*}{Collision Frequency ($f$)} &  Friedman Test & Wilcoxon test with Bonferroni adjustment\\ 
                         &$\chi^2(2) = 17.40, p < .001, W = .347$ &  UIK $>$ FIK ($p=.002\added{, r= .643}$) and MoCap $>$ FIK ($p<.0001\added{, r= .772}$) \\ \hline\addlinespace[0.15cm]
    \multirow{ 2}{*}{Completion Time ($T$)} & One-way within-subject ANOVA  & Pairwise t-test with Bonferroni adjustment \\ 
                         &$F_{2,48} = 4.870, p = .012, \eta^2 = .064$ & MoCap $>$ FIK ($p=.003$) \\ \hline \hline \addlinespace[0.15cm]
                         
  \multicolumn{3}{l}{\textbf{\textit{Pick-and-place Task}}}\\ \hline
    \multirow{ 2}{*}{Volume Per Collision ($v$)} &  Friedman test & -\\
                         &$\chi^2(2) = .72 , p = .698(ns), W = .014$& -\\\hline \addlinespace[0.15cm] 
    \multirow{ 2}{*}{Duration Per Collision ($t$)} &  One-way within-subject ANOVA & Pairwise t-test with Bonferroni adjustment\\
                         &$F_{2,48} = 3.374 , p = .043, \eta^2 = .056 $ & Non-significant\\\hline \addlinespace[0.15cm] 
    \multirow{ 2}{*}{Collision Frequency ($f$)} &  One-way within-subject ANOVA & Pairwise t-test \added{with Bonferroni adjustment}\\
                         &$F_{2,48} = 19.309 , p <.0001,\eta^2 = .209  $ & UIK $>$ FIK ($p<.0001$) and  UIK $>$MoCap ($p<.001$).  \\   \hline\addlinespace[0.15cm]
    \multirow{ 2}{*}{Completion Time ($T$)} &  Friedman Test & Wilcoxon test with Bonferroni adjustment\\ 
                         &$\chi^2(2) = 6.32, p = .042, W = .126 $& UIK $>$ FIK ($p=.017\added{, r= .541}$). \\ \hline \hline \addlinespace[0.15cm]

  \multicolumn{3}{l}{\textbf{\textit{Copy-pose Task}}}\\ \hline
    \multirow{ 2}{*}{Jaccard Distance ($JD$)} &  Friedman test & Wilcoxon test with Bonferroni adjustment.\\
                         &$\chi^2(2) = 24.60 , p<.0001, W = .491$& UIK $>$ FIK ($p=.003\added{, r = .632}$). UIK $>$ MoCap ($p<.0001\added{, r= .848}$). FIK $>$ MoCap ($p=.005\added{, r= .605}$). \\\hline\addlinespace[0.15cm] 
    Mean Per Segment&  Friedman test & Wilcoxon test with Bonferroni adjustment\\
                         \multicolumn{1}{l}{Angle Error ($MPSAE$)} &$\chi^2(2) = 44.20 , p <.0001, W = .885 $ & UIK $>$ FIK ($p<.0001\added{, r= .826}$). UIK $>$ MoCap ($p<.0001\added{, r = .874}$). FIK $>$ MoCap ($p<.0001\added{, r= .864}$). \\\hline\addlinespace[0.25cm] 
                         &  Aligned Rank Transform ANOVA & Tukey's test\\
                         &Animation Fidelity&\\
                         &$F_{2,192} = 179.680 , p <.0001,\eta^2 = .652 $ & UIK $>$ FIK ($p<.0001$). UIK $>$ MoCap ($p<.0001$). FIK $>$ MoCap ($p<.0001$).  \\ \addlinespace[0.15cm]
                         Mean Per Segment &Body Part & \\
                         \multicolumn{1}{l}{Angle Error ($MPPAE$)}& $F_{2,192} = 244.480 , p <.0001,\eta^2 = .718 $ &  Arms $>$ Legs ($p<.0001$) and Arms $>$ Spine ($p<.0001$). \\  \addlinespace[0.15cm]
                  
                         &\multirow{ 8}{*}{$F_{4,192} = 133.460 , p <.0001,\eta^2 = .735 $}& For each Body Part:\\  
                         &&Arms: UIK $>$ FIK ($p<.0001$) and UIK $>$ MoCap ($p<.0001$)\\
                         &Animation Fidelity : Body Part&Legs: UIK$>$FIK ($p<.0001$). UIK$>$MoCap($p<.0001$). FIK$>$MoCap ($p=.003$).\\
                         &&Spine: FIK $>$ UIK ($p<.0001$) and FIK $>$ MoCap ($p<.0001$)\\\addlinespace[0.1cm]
                         &&For each Animation Fidelity condition:\\
                         &&UIK: Arms$>$Legs($p<.0001$). Arms$>$Spine($p<.0001$). Legs$>$Spine($p<.0001$).\\
                         &&FIK: Arms$>$Legs\,($p<.0001$). Arms$>$Spine\,($p<.0001$). Legs$>$Spine\,($p<.0001$).\\
                         &&MoCap: Arms$>$Legs ($p<.0001$) and Arms$>$Spine ($p<.0001$).\\
                         
                         \hline \hline \addlinespace[0.15cm]
  \multicolumn{3}{l}{\textbf{\textit{Questionnaire}}}\\\hline
    \multirow{ 2}{*}{Overall Score} &  \replaced{One-way within-subject ANOVA}{ Friedman test} & \replaced{Pairwise t-test}{ Wilcoxon test} with Bonferroni adjustment\\
                        &\added{$F_{2,48} = 21.033 , p <.0001, \eta^2 = .155$}\\
                        &\deleted{$\chi^2(2) =  18.0, p<.001, W = .491$}  & UIK $<$ FIK (\replaced{$p<.0001$}{$p<.001$}). UIK $<$ MoCap (\replaced{$p<.001$}{$p=.005$}).\\\hline \addlinespace[0.15cm]
    \multirow{ 2}{*}{Agency} &  One-way within-subject ANOVA & Pairwise t-test with Bonferroni adjustment\\
                         &$F_{2,48} = 20.888 , p <.0001, \eta^2 = .168 $ &  UIK $<$ FIK ($p<.0001$) and UIK $<$ MoCap ($p<.001$).\\\hline \addlinespace[0.15cm]
    \multirow{ 2}{*}{Ownership} &  Friedman test & Wilcoxon test with Bonferroni adjustment\\
                         &$\chi^2(2) = 14.5 , p <.001, W = .290 $ & UIK $<$  FIK ($p<.001\added{, r= .771}$). \\ \hline \addlinespace[0.15cm]  

    \multirow{ 2}{*}{Change} &  Friedman test & Wilcoxon test with Bonferroni adjustment\\ 
                         &$\chi^2(2) = 2.06 , p = .358(ns), W = .041 $ & Non-significant \\   
  \bottomrule
  \addlinespace[0.15cm]
  \end{tabu}
  } 
  \caption{Statistic\added{al} results. For task performance data, a higher value implies worse performance. For the questionnaire higher score is better. $W$ value: 0.1-0.3 (small effect), 0.3-0.5 (medium effect) and $\geq$ 0.5 (large effect). $\eta^2$ value: 0.01-0.06 (small effect), 0.06-0.14 (medium effect), $\geq$ 0.14 (large effect). \added{$r$ value: 0.10 - 0.3(small effect),0.30 - 0.5(moderate effect) and $\geq$ 0.5 (large effect).}}
  \label{tab:statistic}
\end{table*}


\begin{table}[tb]
        \scriptsize%
        \centering
    \begin{tabu}{%
	l%
	*{4}{r}%
	}
     & \multicolumn{1}{c}{$v$} & \multicolumn{1}{c}{$t$} & \multicolumn{1}{c}{$f$} & \multicolumn{1}{c}{$T$} \\ 
    
    \toprule
    \multicolumn{2}{l}{\textbf{\textit{Spike-over-spikes Task}}}\\
     \textbf{UIK} & $101.0(61.7)$ & $0.060(0.067)$ & $0.189(0.189) $ & $102.0(25.6) $\\
     \textbf{FIK}& $\mathbf{86.2(81.3)}$ & $\mathbf{0.044(0.038)} $ & $\mathbf{0.099(0.125)} $ & $\mathbf{95.3(20.7)} $ \\
      
      \textbf{MoCap}&$126.0(34.4)$& $0.068(0.036) $ & $0.361(0.377) $ & $110.0(23.6) $ \\\addlinespace[0.15cm]
        
     \multicolumn{2}{l}{\textbf{\textit{Pick-and-place Task}}}\\
    \textbf{UIK} & $187.0(57.2) $ & $0.271(0.090) $ & $0.516(0.305) $ & $101.0(38.0)$ \\
     
      \textbf{FIK}& $183.0(81.9) $ & $0.280( 0.085) $ & $\mathbf{0.268(0.178)} $ & $\mathbf{78.1(27.4)}$\\
      
      \textbf{MoCap} & $\mathbf{171.0(65.9)} $ & $\mathbf{0.231(0.091)} $ & $0.283(0.166) $ & $80.6(19.0)$\\
      
    \bottomrule
    \end{tabu}
    \caption{Mean and standard deviation for metrics of step-over-spikes task and pick-and-place task.}
    \label{tab:dynamic}
\end{table}

     
      

\begin{figure*}[tb]
 \centering 
 \includegraphics[width=0.9\textwidth]{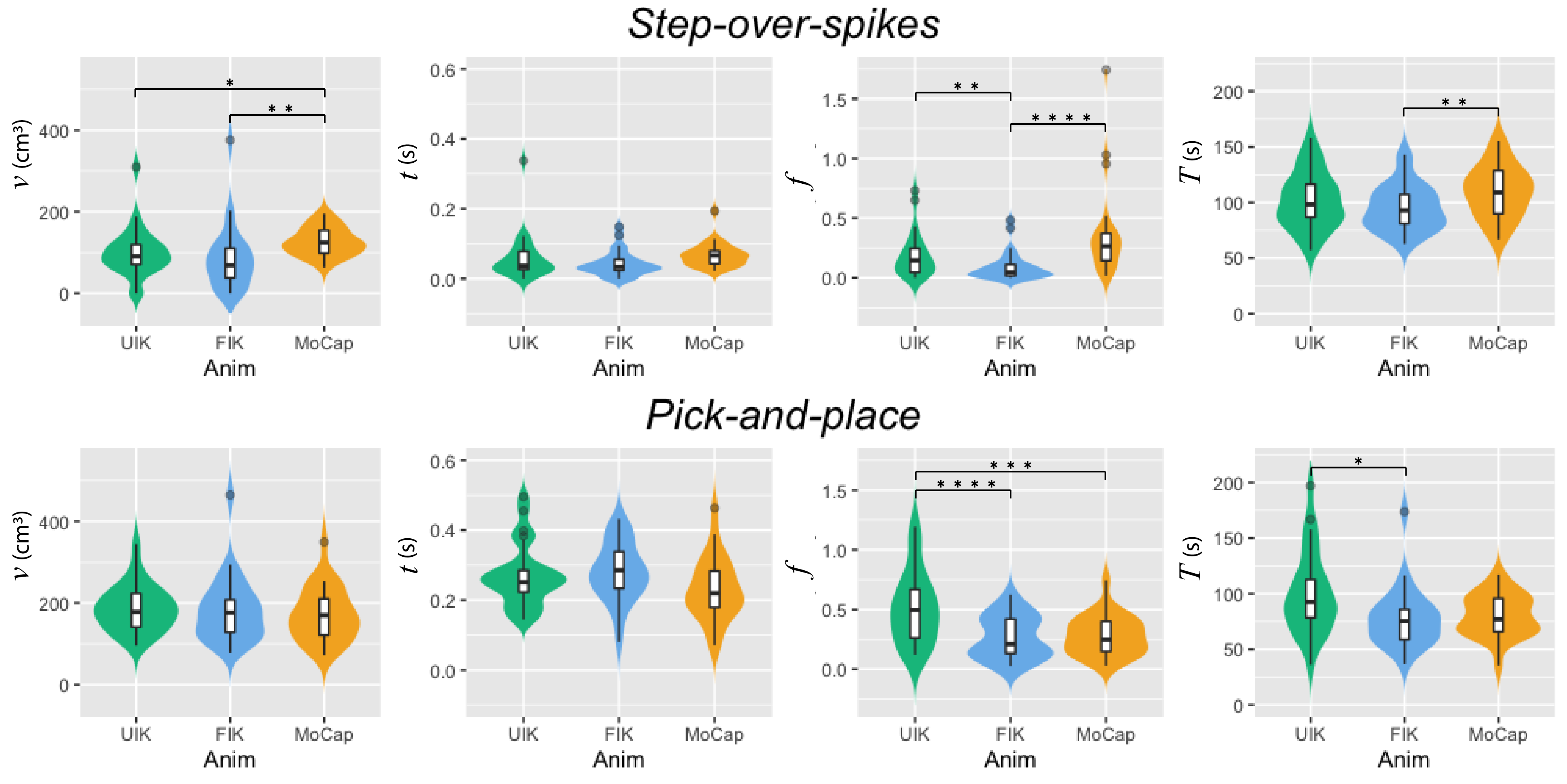}
 \caption{Violin plots for metrics of the step-over-spikes and pick-and-place tasks, showing results for collisions and task completion time. Asterisks represent the significance level: * (p $<$ .05), ** (p $<$ .01), *** (p $<$ .001), **** (p $<$ .0001).}
 \label{fig:Dynamic_overview}
\end{figure*}

\subsection{User performance on interaction tasks} 
We first present the results of user performance on the tasks that involved a direct interaction with the \replaced{VE}{virtual environment}. Table~\ref{tab:dynamic}\deleted{~and~\ref{tab:UpperData}} show\added{s} the mean (M) and standard deviation (SD) of all the metrics of the step-over-spikes and pick-and-place tasks. \replaced{Fig.}{Figure}~\ref{fig:Dynamic_overview} show\added{s} the violin plots of the metrics of both tasks. 

Shapiro-Wilk tests showed significant departures from normality for all three measures of the step-over-spikes task. Therefore, non-parametric within-subjects Friedman tests were used and they revealed significant differences for all metrics between animation fidelity conditions. Animation fidelity significantly affected volume per collision and collision frequency but not duration per collision. Table~\ref{tab:statistic} includes a summary of $\chi^2(2)$, p-values and effect sizes calculated for these metrics. \replaced{Pairwise post-hoc tests (Wilcoxon signed-rank tests) showed that}{Post-hoc Wilcoxon signed-rank tests were conducted to check the significance of pairwise comparisons. In pairwise post-hoc tests,} MoCap \deleted{condition} had significantly higher values than FIK for all metrics except duration per collision, and a significantly higher value than UIK for collision frequency.
\replaced{It also showed UIK had significantly higher collision frequency than FIK}{A significant difference was also found between FIK and UIK in collision frequency}.

For the pick-and-place task, Shapiro-Wilk tests showed that volume per collision and completion time data violated \added{the} normality assumption (p$<$.05)\added{,} while the other two metrics did not. Therefore, Friedman tests and post-hoc Wilcoxon signed-rank tests were conducted for volume per collision and completion time\added{,} while one-way within-subject ANOVAs and pairwise t-tests were conducted for the others. The results revealed a significant effect of animation fidelity on duration per collision and collision frequency. Post-hoc \deleted{pairwise} tests showed that UIK had significantly higher collision frequency than FIK and MoCap, and a longer completion time than FIK. 

Therefore, hypothesis \textbf{[H1]} was validated by these results of interactions tasks performed for both the upper body and lower body. \added{We further analyze these results in Section~\ref{sec:disc}.}

\subsection{User performance on pose-related tasks} 
\begin{figure*}[tb]
 \centering 
 \includegraphics[width=0.9\linewidth]{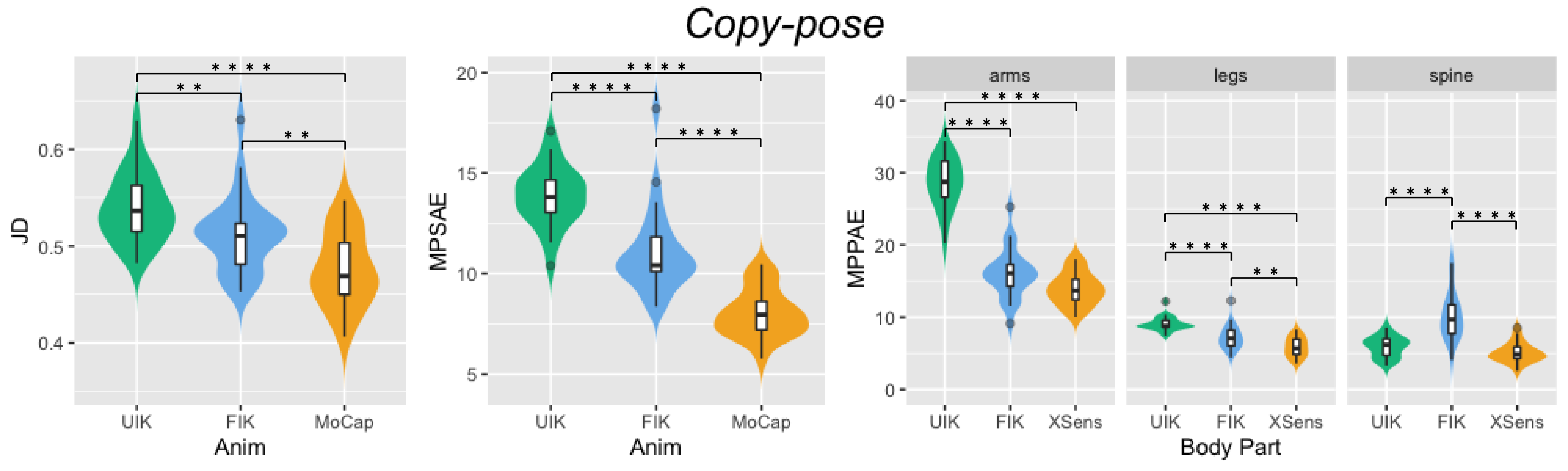}
 \caption{Violin plots of metrics obtained for the copy-pose task. Asterisks represent the significance level: * (p $<$ .05), ** (p $<$ .01), ***~(p~$<$~.001), **** (p $<$ .0001).}
 \label{fig:CP_overview}
\end{figure*}

We summarize the \replaced{M}{mean} and SD for all metrics of the copy-pose task in Table \ref{tab:copy-pose} and present the corresponding violin plots in \replaced{Fig.}{Figure}~\ref{fig:CP_overview}. Shapiro-Wilk tests showed \added{both JD and MPSAE data had} \added{a} non-significant departure\added{s} from normality (p $<$ .05)\deleted{for JD and MPSAE data}. Friedman tests were \added{thus} conducted \added{for both metrics} and revealed significant differences among the three animation fidelity conditions with medium to large effect sizes\deleted{, for both metrics}. \added{Pairwise} Wilcoxon test\added{s} with Bonferroni p-value adjustment \deleted{was conducted for pairwise comparisons and showed} demonstrated significant differences in all pairs of conditions. For both metrics, UIK had significantly higher error values than FIK and MoCap, and FIK had significantly higher errors than MoCap. 

For MPPAE, we used a two-way repeated measures Aligned Rank Transform (ART) ANOVA after asserting the normality with a Shapiro-Wilk test (p $<$ .05). The result revealed a significant main effect of animation fidelity and body part on MPPAE. It also showed a significant interaction effect between animation fidelity and body part. First, the post-hoc Tukey’s tests demonstrated that, for all animation fidelity conditions, MPPAE was significantly higher for arms than for legs and spine. Next, when comparing the MPPAE for each body part, Tukey's tests showed that, for arms, the MPPAE was significantly higher for UIK than for the other conditions. For legs, UIK had significantly higher MPPAE than FIK and MoCap, and FIK had significantly higher MPPAE than MoCap. For the spine, FIK had significantly higher MPPAE than other conditions.

To summarize, these results validated our hypothesis \textbf{[H2]} in the sense that the pose errors were significantly lower when using MoCap than IK solutions.
\begin{table}[tb]
        \scriptsize%
        \centering
    \begin{tabu}{%
	l%
	*{4}{r}%
	}
     & $JD$ & $MPSAE$ & \multicolumn{2}{c}{$MPPAE$}\\ 
    
    \toprule
     \multirow{ 3}{*}{\textbf{UIK}}
     & & &Arms & $28.6 (3.45)$\\
     & $0.539 (0.035) $& $13.90 (1.51)$ & Legs& $9.09 (0.98)$\\ 
     & & &Spine& $5.90 (1.43)$\\
     
      \multirow{ 3}{*}{\textbf{FIK}}
      & & &Arms & $16.1 (3.45)$\\
      & $0.512 (0.040) $ & $11.10 (2.10) $& Legs& $7.22 (1.67)$\\
      & & &Spine& $10.1 (3.26)$\\
      
      \multirow{ 3}{*}{\textbf{MoCap}}
      & & &Arms & $\mathbf{13.9 (2.23)}$\\
      & $\mathbf{0.476 (0.038) }$& $\mathbf{8.03 (1.19) }$& Legs& $\mathbf{5.85 (1.33)}$\\
      & & &Spine& $\mathbf{5.08 (1.38)}$\\

    \bottomrule  
    \end{tabu}
    \caption{Mean and standard deviation for metrics of the copy-pose task.}
    \label{tab:copy-pose}
\end{table}

\subsection{Sense of Embodiment}

\begin{figure*}[tb]
 \centering 
 \includegraphics[width=0.9\linewidth]{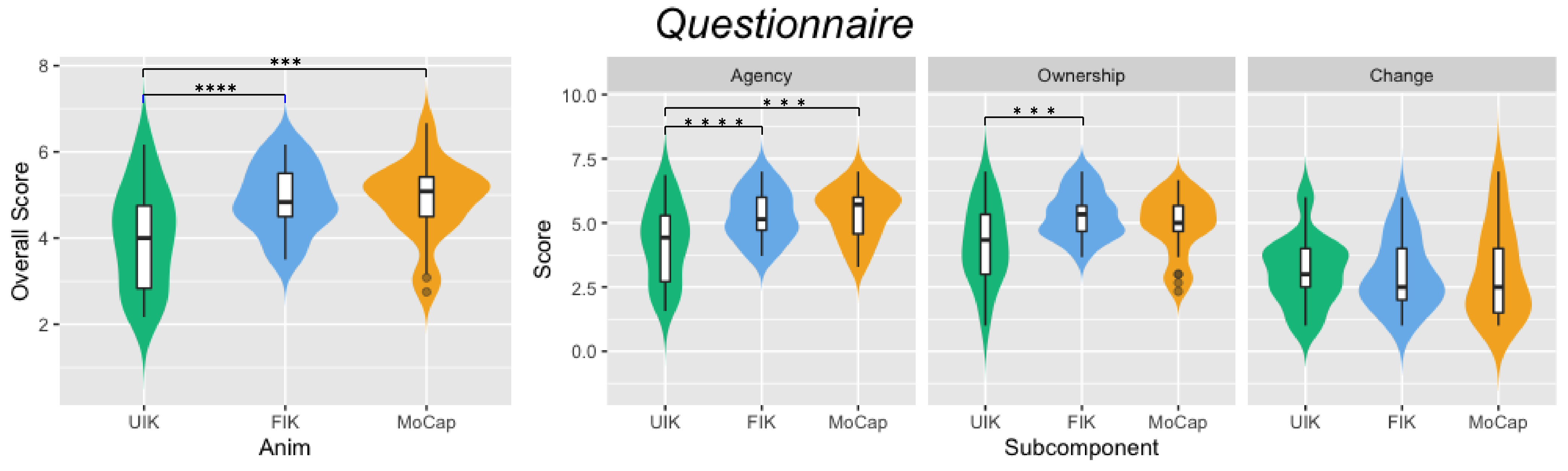}
 \caption{Violin plots for overall score of the SoE and scores for agency, ownership and change individually. Asterisks represent the significance level: * (p $<$ .05), ** (p $<$ .01), *** (p $<$ .001), **** (p $<$ .0001). }
 \label{fig:q}
\end{figure*}

Table \ref{tab:QuestionData} shows the M and SD of the overall score of the SoE and subcomponent scores for \textit{agency}, \textit{ownership} and \textit{change}. The violin plots for these scores can be found in \replaced{Fig.}{Figure}~\ref{fig:q}. \replaced{A one-way within-subject ANOVA showed a significant effect of animation fidelity on overall score of the SoE. The post-hoc tests (pairwise t-test) showed that the SoE score for UIK was worse than both FIK and MoCap.}{For the overall score of the SoE, the normality assumption was violated as assessed by a Shapiro-Wilk test (p $<$ .05). A Friedman test was conducted and showed a significant main effect of animation fidelity on the overall score of the SoE. The post-hoc tests (Wilcoxon test with Bonferroni p-value adjustment) showed that the SoE score for UIK was worse than both FIK and MoCap.}

We analyzed the average score of \textit{agency} questions, Q1 - Q7, with a \replaced{one-way}{One-way} within-subject ANOVA (see \replaced{Fig.}{Figure}~\ref{tab:statistic} for test values). The result showed a significant effect of animation fidelity on \textit{agency} score. The post-hoc test\added{s} (pairwise t-test) showed that users reported the SoA in UIK significantly lower than FIK and MoCap.

Since \added{a} Shapiro-Wilk test showed a non-significant departure from normality, a Friedman test was conducted for the average score of \textit{ownership} questions, Q8 - Q10. The result showed a significant effect of animation conditions on ownership. The post-hoc test (Wilcoxon test with Bonferroni p-value adjustment) showed UIK had a significantly lower \textit{ownership} score than FIK. 

The same set of tests as \textit{ownership} were conducted for the average score of \textit{change} questions, Q11 and Q12. A Friedman test showed no significant effect of animation conditions on \textit{change}. post-hoc test\added{s} showed no significant difference on \textit{change} in all condition pairs. Overall, these results validated our hypothesis \textbf{[H3]}.
\begin{table}[tb]

    \scriptsize%
        \centering
    \begin{tabu}{%
	l%
	*{7}{l}%
	}
     & $Overall$ & $Agency$ & $Ownership$ & $Change$  \\
    \toprule
     \textbf{UIK}&\replaced{$4.03(1.18)$}{$4.33(1.32)$}  &$4.22(1.51)$&$4.19(1.54) $ &\replaced{$3.1(1.34)$}{$4.90 (1.34)$} \\
      \textbf{FIK}&\replaced{$\mathbf{4.91 (0.774)}$}{$\mathbf{5.20 (0.81)}$} &$5.29 (0.97)$&$\mathbf{5.32(0.92) }$ &\replaced{$\mathbf{2.98(1.35)}$}{$\mathbf{5.02(1.35)}$} \\
      \textbf{MoCap}&\replaced{$4.87(0.919)$}{$5.18(0.89$)} &$\mathbf{5.37(0.99)}$&$4.91(1.20) $ &\replaced{$3.08(1.79)$}{$4.92(1.79)$} \\
    \bottomrule  
    \end{tabu}
        \caption{Mean and standard deviation for the overall score of the SoE and scores of subcomponents.}
    \label{tab:QuestionData}

\end{table}

\section{Discussion}
\label{sec:disc}
\subsection{Accuracy of body pose vs. end-effector positioning}

As expected, a motion capture suit is able to capture most of the human motion and\deleted{to} accurately represent poses, as opposed to applying IK using only a few trackers as end-effectors. We quantitatively assessed this with the copy-pose task and found that the MoCap method performed significantly better than UIK and FIK for all metrics. Poses with MoCap were best aligned with the given poses and also when analyzing each body segment independently.

\begin{figure}
 \centering 
 \includegraphics[width=0.9\linewidth]{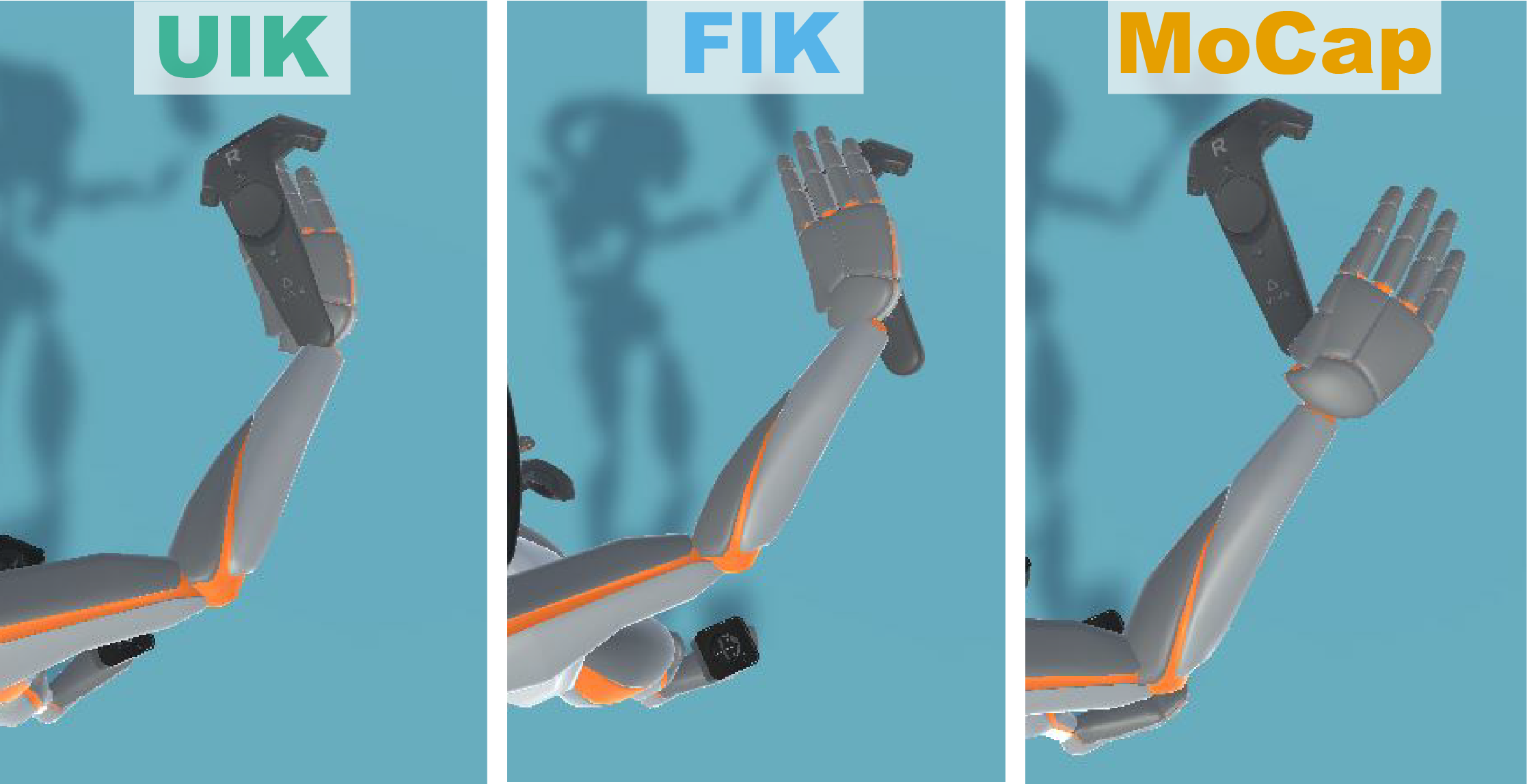}
 \caption{End-effectors positioning with respect to controllers for
the different animation conditions.}
 \label{fig:supply}
 \vspace{-4mm}
\end{figure}

Therefore, we would expect MoCap to perform better in other tasks due to the high-quality tracking of poses. However, we found that high-quality poses do not improve task performance when tasks are not directly related to the pose, and instead require direct interactions with the \replaced{VE}{virtual environment}. One possible explanation is that the positional drift from inertial systems results in the end-effectors moving away from their actual position. When this happens, the user's hands and feet are no longer co-located with their virtual representations, thus introducing inconsistencies \added{(see Fig. \ref{fig:supply})}. \added{The higher latency of MoCap may have also contributed to these performance differences.}

In the step-over-spikes task, MoCap was significantly worse than FIK in volume per collision, collision frequency and completion time. MoCap was significantly worse than UIK in volume per collision. We believe that for this task, having an accurate positioning of the feet (no drift) made users feel more confident and reliable when positioning the feet on the ground to avoid \deleted{the} spikes. Both FIK and UIK achieved good \replaced{foot}{feet} positioning because IK solvers enforce\added{d} the position of the feet to be the same as the trackers. In contrast, since MoCap is an IMU-based motion capture system, it does not have precise global positioning of the joints.

To lessen the positional drift issue, we moved the MoCap avatar to match the position of the VR pelvis tracker. This improves the overall co-location between the user and its avatar, but it may increase foot sliding. For instance, when one leg is acting as a supporting leg on the ground as the user\added{'s} pelvis moves, if the pelvis of the MoCap animated avatar is forced to follow the HTC pelvis tracker, it makes the foot slide on the ground and increases the risk of collision with obstacles. To minimize this problem\added{,} we implemented a foot lock algorithm. This alleviated foot sliding but not the global position accuracy of the feet.   

Overall, if the task requires accurate foot placement, it may be necessary to include foot trackers to position them accurately in the \replaced{VE}{virtual environment}, while correctly posing all joints may not be necessary.



\subsection{Upper body animation for accurate interactions with the environment}

In the pick-and-place task, UIK performed significantly worse than MoCap and FIK in terms of collision frequency. However, we found MoCap and FIK to perform similarly. This is consistent with the results of the MPPAE in the copy-pose task, for which UIK also performed worse than MoCap and FIK due to incorrect elbow positioning. For the pick-and-place task, users had to correctly position the arm to reach the goal without touching the obstacles. The incorrect elbow positioning in UIK made the task more complicated, and thus more prone to collisions. We also found that users took significantly \replaced{longer to finish}{more time to complete} the task with UIK than \added{with} FIK. 

When comparing FIK and MoCap, our results suggest that the additional tracking data for the elbows in MoCap did not help the participants achieve better performance in terms of collision avoidance in the pick-and-place task, and also in the arm part of pose replication in the copy-pose task. Even though MoCap provides a more accurate \replaced{elbow position}{position of the elbow}, we believe that the inaccurate end-effector positions lead to more collisions with the obstacles. Another explanation may be due to the latency of the MoCap. A few participants commented that their virtual arms were less responsive when using MoCap while performing the pick-and-place task. \added{As Waltemate et al. \cite{Waltemate} stated, when latency increases above $75\,ms$, user's motor performance in VR tends to decline.}

Even if FIK provides less accurate poses for the elbows, its responsiveness and end-effector accuracy compensate for this. Participants can quickly avoid obstacles by adjusting the controllers' position. The result is consistent with the work by Parger et al. \cite{parger_HumanUpperBodyInverse_2018}. 

\subsection{Performance differences between arms and legs}
The results of the MPPAE in the copy-pose task suggest that the arm poses were less precise than the leg poses. The angle error was larger in the arms than in the legs for all conditions. One possible explanation is that the range of movements a human person can do with their upper body is wider than with the lower body. We also studied whether users noticed the tracking inaccuracy by comparing the scores given in questions related to arms (Q2-Q4) and legs (Q5-Q6). The score for arms ($M = 4.60$, $SD = 1.19$) \replaced{was}{is} statistically ($p < .0001$) lower than legs ($M = 5.41$, $SD = 1.05$) when performing a t-test. When performing a two-way ANOVA, adding the animation fidelity as a condition, we found no statistical difference between the scores given to the arms questions between FIK and MoCap. 

\replaced{The participant-reported differences in responsiveness between FIK and MoCap for arm movement were not observed for the legs during the step-over-spikes task.}{The differences in responsiveness between FIK and MoCap that were reported by participants for the arm movement, were not observed for the legs when performing the step-over-spikes task.}


Based on the result above, we recommend focusing on the upper body when animating a self-avatar since it seems necessary to have higher-quality animations for arms. Lower-quality animation may be enough for the legs. Therefore, as some works have suggested \cite{ponton2022mmvr}, it may not be necessary to include all tracker devices for the lower body when the task does not require accurate foot placement. 



\subsection{High Sense of Agency can be achieved with a small set of tracking devices}

The questionnaire data showed no statistically significant differences between FIK and MoCap. However, as mentioned before, MoCap achieved better results ($JD$ and $MPSAE$) than FIK and UIK in the copy-pose task. It suggests that \added{the} SoA is not only related to the pose, but also \added{to} the interaction with the \replaced{VE}{virtual environment}, e.g., we found \added{that} in the pick-and-place task MoCap did not achieve the best results.

In other words, our results suggest that one can feel the same level of control over self-avatars animated by a high-end motion capture suit with 17 IMUs or a small set of tracking devices (one HMD, two controllers, and three trackers) and a high-quality IK solution. This finding is consistent with Galvan Debarba et al. \cite{galvandebarba_PlausibilityVirtualBody_2020} that suggested a total of 8 trackers were enough to achieve the same plausibility illusion as an optical-based motion capture system with 53 retro-reflective markers. Goncalves et al. \cite{goncalves_Evaluationimpactdifferent_2022} suggested that increasing tracking points, from 3 to 6, does not significantly improve the \replaced{SoA}{Sense of Agency}.

More research is needed to understand how to improve the SoA, given that a higher number of trackers (MoCap) did not always improve the agency scores when compared to a full-body IK such as FIK. Other factors such as end-effectors position accuracy, latency or animation smoothness may affect the users' perception.

\added{It would also have been interesting to randomize the task order so that we could have analyzed whether the results of the SoE were affected by which was the last task being experienced by the participant. However, by looking at the results, we observe that the step-over-spike task (the first task) had FIK giving better quantitative results, the pick-and-place task (the second task) had similar performance for FIK and Mocap, and in the copy-pose task (the last task) MoCap had the best results. Even though the last task had better performance for Mocap, the embodiment questionnaires showed similar results for FIK and MoCap (not statistically significant) which may indicate that the questionnaire did gather the overall experience.} 

\section{Conclusions and Future Work}

We conducted a user study to examine the impact of the avatar's animation fidelity on user performance and the SoA. Our results suggest that the IMU-based motion capture system performed better than IK solutions for applications that require pose accuracy. However, IK solutions outperform IMU-based motion capture systems when directly interacting with the \replaced{VE}{virtual environment}. In these cases, accurate end-effector placement \added{and low latency} may be more critical than exact pose matching due to proprioception. Our study also suggests that a high-end IK solution with sparse input (6 trackers) can achieve similar levels of the \replaced{SoA}{Sense of Agency} as an IMU-based motion capture with dense input (17 trackers). We believe these results give insight into how animation fidelity affects user performance and perception, providing future research directions toward improving self-avatar animation fidelity in VR. \added{Our work also highlights the limitations of current technology to achieve correct self-avatar animation (such as latency, end-effectors and body pose inaccuracy), and thus motivates future research to overcome these issues.}

\added{
A limitation of our experiment is that the robotic avatar did not accurately match the shape of the participant. Since the avatar's limbs were much thinner than the participants' ones, and because they used hand-held controllers, self-contacts suggested by some copy-pose targets were not reproduced by the avatar (regardless of the condition). In fact, no participant referred to this issue. Further studies are required to study the role of animation fidelity and self-contact \cite{bovet2018critical} when the avatar accurately matches the user's shape.
}

For future research, we would like to investigate whether participants could perform better using an optical motion capture system, providing both accurate pose and global position. This new condition will allow the decoupling of the positional drift issue from the accuracy of the body pose, allowing for a more in-depth study of the perceptual results. \replaced{We believe future studies that integrate hand tracking like RotoWrist \cite{Parizi} or data-driven methods for self-avatar animation would be valuable to provide more insight into how animation fidelity impacts the SoE and user performance in VR.}{Another interesting direction would be to study more recent data-driven solutions reconstructing full-body motion from sparse tracking data.}

\acknowledgments{
This project has received funding from the European Union’s Horizon 2020 research and innovation programme under the Marie Skłodowska-Curie grant agreement No. 860768 (CLIPE project) and from MCIN/AEI/10.13039/501100011033/FEDER, UE (PID2021-122136OB-C21). Jose Luis Ponton was also funded by the Spanish Ministry of Universities (FPU21/01927).
}

\clearpage 
\bibliographystyle{abbrv-doi}

\bibliography{template}
\end{document}